\documentclass[final,5p,times,twocolumn,authoryear]{elsarticle}

\usepackage{amssymb}
\usepackage{amsmath}
\usepackage{graphicx}
\usepackage{xcolor}
\usepackage{dblfloatfix}
\usepackage{placeins}

\newcommand{\pccm}{\ensuremath{\mathrm{pc\,cm^{-3}}}}
\newcommand{\dtdnu}{\ensuremath{\mathrm{d}t/\mathrm{d}\nu}}
\newcommand{\tauemp}{\ensuremath{\tau_{\rm emp}}}
\newcommand{\tausc}{\ensuremath{\tau_{\rm s}}}
\newcommand{\sigint}{\ensuremath{\sigma_{\rm int}}}

\newcommand{\reffreq}{\ensuremath{583~\mathrm{MHz}}}

\journal{New Astronomy}

\begin{document}

\begin{frontmatter}

\title{A frequency-resolved measurement of the scattering law in
FRB~20210415C}

\author[first]{Victor Pop}
\ead{popv@mcmaster.ca}
\author[second,third]{Mohammed A. Chamma}
\author[fourth]{Aishwarya Kumar}

\author[first]{Fereshteh Rajabi}

\affiliation[first]{
    organization={Department of Physics and Astronomy, McMaster University},
    addressline={1280 Main Street West},
    city={Hamilton},
    postcode={L8S 4M1},
    state={Ontario},
    country={Canada}
}

\affiliation[second]{
    organization={Department of Physics, McGill University},
    addressline={3600 rue University},
    city={Montréal},
    postcode={H3A 2T8},
    state={Québec},
    country={Canada}
}

\affiliation[third]{
    organization={Trottier Space Institute, McGill University},
    addressline={3550 rue University},
    city={Montréal},
    postcode={H3A 2A7},
    state={Québec},
    country={Canada}
}

\affiliation[fourth]{
    organization={Department of Physics and Astronomy, The University of Western Ontario},
    addressline={1151 Richmond Street},
    city={London},
    postcode={N6A 3K7},
    state={Ontario},
    country={Canada}
}

\begin{abstract}
Fast radio bursts (FRBs) show diverse temporal morphologies, shaped
both by their intrinsic emission and by propagation effects such as
residual dispersion and multipath scattering. Because several of these
effects can broaden a burst asymmetrically, an asymmetric profile does
not by itself establish multipath scattering. We measure the frequency dependence of the temporal tail in FRB~20210415C, a bright burst from the repeating source FRB~20210219C. The burst is detected across approximately 400--800~MHz. Frequency-resolved fits yield an empirical
exponential-tail timescale, $\tauemp$, that follows
$\tauemp\propto\nu^\alpha$. Two complementary treatments of the
underlying pulse width give $\alpha=-3.95\pm0.07$ and
$-4.14\pm0.08$. Both results are consistent with the commonly adopted
$\nu^{-4}$ scattering law. The inferred scattering time is
approximately 8.0~ms at \reffreq, exceeding Galactic foreground
estimates by a factor of at least $4\times10^2$. This favours a
dominant contribution beyond the Milky Way, although the data do not
localise the scattering screen. FRB~20210415C therefore provides a direct measurement of a scattering
law commonly imposed in catalogue analyses.
\end{abstract}

\begin{keyword}
fast radio bursts \sep radio transients \sep interstellar scattering
\sep pulse broadening \sep methods: data analysis
\end{keyword}

\end{frontmatter}

\section{Introduction}
\label{sec:introduction}

Fast radio bursts (FRBs) are bright, typically millisecond-duration
radio transients \citep{lorimer2007,petroff2022,zhang2023}, although
bursts lasting only microseconds have now been detected
\citep{snelders2023}. Their signals propagate through
ionised plasma in their local environments, host galaxies, intervening
media, and the Milky Way \citep{petroff2022}. Small-scale electron-density inhomogeneities redirect the radiation
along multiple paths. The resulting differences in path length produce
arrival-time delays and an asymmetric temporal tail. In the idealised, infinitely extended thin-screen approximation, the
pulse-broadening function is commonly represented by a one-sided
exponential
\citep{cronyn1970,williamson1972,rickett1977,
geyerkarastergiou2016}, with characteristic timescale

\begin{equation}
    \tausc(\nu)
    =
    \tau_{\rm ref}
    \left(
        \frac{\nu}{\nu_{\rm ref}}
    \right)^{\alpha}.
    \label{eq:scattering_law}
\end{equation}
Here, $\tausc(\nu)$ is the scattering time at observing frequency
$\nu$, $\tau_{\rm ref}$ is its value at the reference frequency
$\nu_{\rm ref}$, and $\alpha$ is the scattering-frequency index. With
this convention, $\alpha<0$ corresponds to stronger broadening at
lower frequencies. In idealised thin-screen models, Gaussian or
single-scale density fluctuations give $\alpha\simeq-4$, whereas an
ideal Kolmogorov spectrum gives $\alpha=-4.4$ under standard
strong-scattering assumptions
\citep{rickett1977,leejokipii1975,geyerkarastergiou2016}. Extended or otherwise
non-ideal scattering media can modify both the pulse-broadening
function and its effective frequency dependence
\citep{williamson1974,geyerkarastergiou2016}. An index near $-4$ is
therefore consistent with scattering but does not uniquely determine
the screen geometry or fluctuation spectrum.

A trailing profile alone, however, does not establish multipath
scattering. Unresolved burst components, intrinsic profile evolution,
residual dispersion, instrumental resolution, and the summation of
drifting emission across a finite bandwidth can also produce broad or
asymmetric profiles. A full-band exponentially modified Gaussian
(EMG) fit can therefore establish that an asymmetric component is
required, but not determine its physical origin. A more discriminating
test is to resolve the empirical tail across frequency and determine
whether its timescale follows the steep, monotonic dependence expected
from scattering. Such a measurement must also account for covariance
between the exponential-tail timescale and the width of the underlying
pulse component.

Direct measurements of the scattering-frequency index remain difficult
for individual FRBs. For FRB~110220, \citet{Thornton2013} measured
$W\propto\nu^{-4.0\pm0.4}$, consistent with scattering. More recently, \citet{gopinath2024} fitted the scattering-frequency index for eight bursts from FRB~20180916B, obtaining
$-4.5\leq\alpha\leq-1.7$ with $1\sigma$ uncertainties of 0.3--2.3. They therefore referenced the reported scattering times to 150~MHz assuming $\alpha=-4$, incorporating the effect of plausible deviations from this value into the uncertainties. The Second CHIME/FRB Catalog likewise adopts $\alpha=-4$ when fitting the broadening amplitude
\citep{chime2026}; it therefore tests for a prescribed $\nu^{-4}$ tail
rather than measuring the exponent independently. A sufficiently bright
burst observed over a broad fractional bandwidth can instead constrain
the scattering law directly.

Here we analyse FRB~20210415C, an individual burst from the repeating
source FRB~20210219C in the Second CHIME/FRB Catalog
\citep{chime2026}. We first test whether its frequency-integrated
profile requires an exponential tail, then resolve that tail across
the CHIME band after correcting for frequency drift. We leave
$\alpha$ free and use complementary treatments of the
Gaussian-component width to test whether the inferred frequency
dependence is affected by width--tail covariance. The primary aim is to measure the scattering law directly rather than
infer a broadening amplitude after imposing $\alpha=-4$. As a
complementary propagation diagnostic, we also search for Galactic
diffractive scintillation, which can provide additional information
about the line-of-sight scattering geometry.

The remainder of this paper is organised as follows.
Section~\ref{sec:analysis} describes the full-band profile modelling,
frequency-resolved tail measurement, and scintillation search.
Section~\ref{sec:results} presents the measured scattering law,
Gaussian-width evolution, and scintillation constraints, while
Section~\ref{sec:discussion} discusses their physical interpretation
and limitations. Technical details of the profile fitting and model
selection are given in~\ref{app:fitting}, the robustness
tests in~\ref{app:robustness}, and the
spectral-autocorrelation and scintillation injection--recovery
analyses in~\ref{app:scint}.

\section{Observation and analysis}
\label{sec:analysis}

The analysis proceeds in three parts. We first determine whether the
frequency-integrated profile requires a one-sided tail, then measure
the frequency dependence of that tail, and finally search the burst
spectrum for diffractive scintillation as a complementary propagation
diagnostic.

\subsection{Data and profile model}
\label{sec:data}

\begin{figure*}[!t]
\centering
\includegraphics[width=\textwidth]
{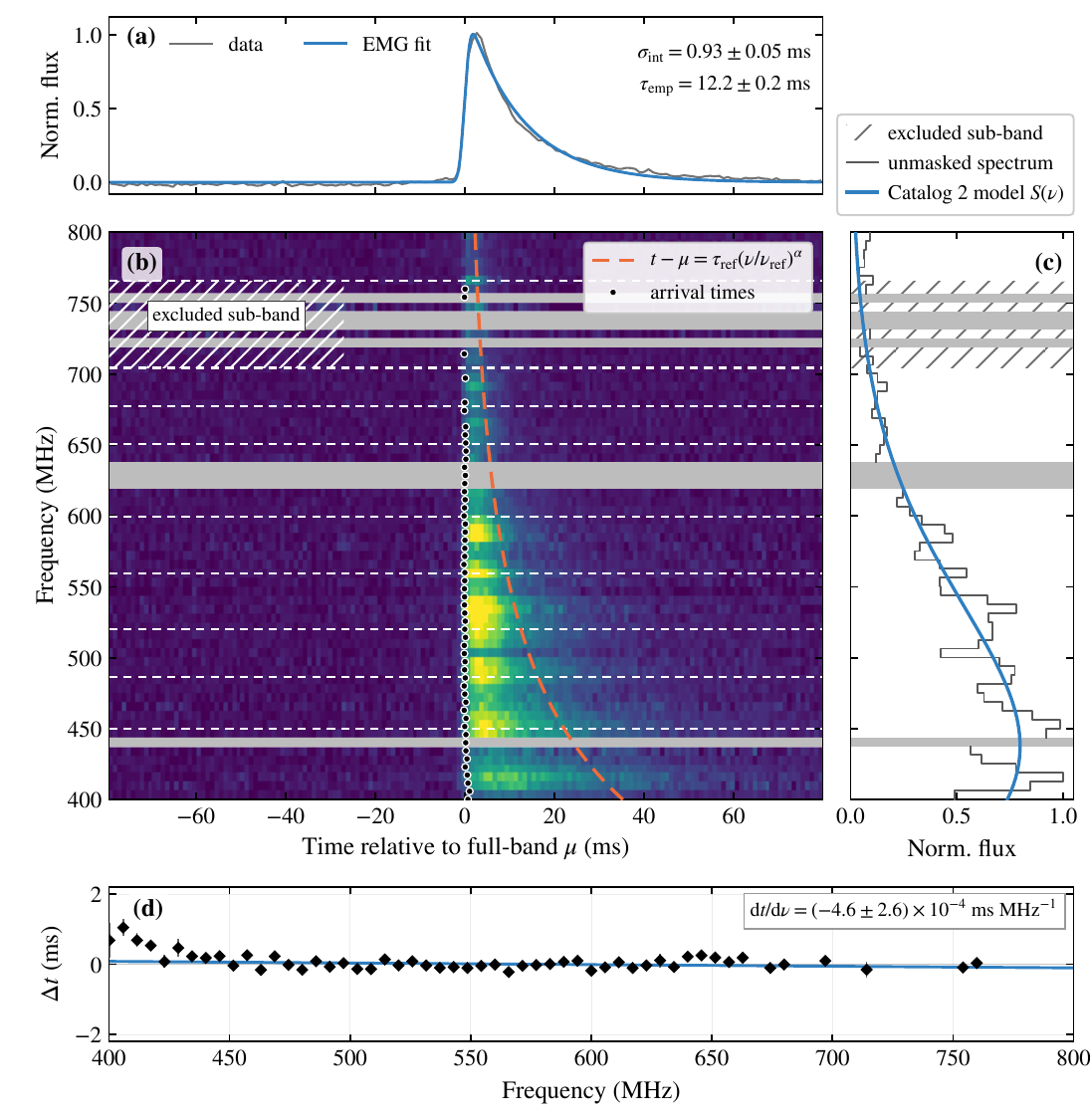}
\caption{
Overview of FRB~20210415C and the quantities used in the analysis.
\emph{(a)} Frequency-integrated Stokes-$I$ profile (grey) with the
retained bin-integrated EMG fit (blue).
\emph{(b)} Dedispersed dynamic spectrum, averaged into 6.25-MHz display
channels; grey rows are masked channels. Time is measured from the
centre $\mu$ of the full-band Gaussian component. Black points mark the
53 per-channel arrival times, defined as the leading-edge half-maximum
time of each channel's fitted EMG profile. Horizontal dashed lines mark
the boundaries of the nine sub-bands; channels above the uppermost line
are not used in the sub-band analysis, and the hatched
highest-frequency sub-band is excluded from the primary
tail--frequency fit but retained for the all-nine-sub-band
Gaussian-width diagnostic. The dashed curve traces
$t-\mu=\tau_{\rm ref}(\nu/\nu_{\rm ref})^{\alpha}$ from the free-width
fit, corresponding to one fitted e-folding time of the tail.
\emph{(c)} Time-integrated spectrum of the unmasked channels (grey
line); the grey bands mark the masked intervals in panel~(b). The blue
curve is the running-power-law spectral model from the Second
CHIME/FRB Catalog, $S(\nu)$, used to define the flux-weighted sub-band
frequencies and the injected spectra. The hatched band marks the
highest-frequency sub-band; its unmasked channels are shown for
completeness but do not enter the eight-sub-band tail--frequency fit.
\emph{(d)} Per-channel arrival times relative to the fitted linear
relation at $\nu_{\rm ref}$, with fitted slope
$\dtdnu=(-4.6\pm2.6)\times10^{-4}~\mathrm{ms\,MHz^{-1}}$.
}
\label{fig:waterfall}
\end{figure*}

The public Catalog~2 Stokes-$I$ data product has a native time
resolution of 0.983~ms and is incoherently dedispersed using
$\mathrm{DM}=526.3~\pccm$ \citep{chime2026}. The detectable emission
spans most of the 400--800-MHz CHIME band; the frequency-resolved
analysis below uses the interval 400.2--765.6~MHz.
Figure~\ref{fig:waterfall} provides an overview of the burst:
panel~(a) shows the frequency-integrated profile and its pronounced
trailing tail, while panel~(b) shows the dedispersed dynamic spectrum
used for the frequency-resolved analysis.

To characterise the profile asymmetry, we model the
frequency-integrated Stokes-$I$ profile, $I(t)$, as an exponentially
modified Gaussian: a Gaussian component convolved with a one-sided
exponential response,

\begin{equation}
\begin{aligned}
I(t)
&=
A\left[
G(t;\mu,\sigint)
*
E(t;\tauemp)
\right]+C,\\
E(t;\tauemp)
&=
\frac{H(t)}{\tauemp}
\exp\left(-\frac{t}{\tauemp}\right).
\end{aligned}
\label{eq:emg}
\end{equation}
Here, $G(t;\mu,\sigint)$ is a Gaussian with centre $\mu$ and standard
deviation $\sigint$, while $E(t;\tauemp)$ is an exponential response
with e-folding time $\tauemp$. The Heaviside function $H(t)$ sets the
response to zero for negative delays, so that the exponential produces
only a trailing component. The symbol $*$ denotes convolution, while
$A$ and $C$ set the profile amplitude and constant baseline,
respectively.

Because unresolved temporal structure and instrumental broadening may
contribute to the Gaussian component, $\sigint$ need not represent the
fundamental emission duration. We retain the empirical notation
$\tauemp$ until its frequency dependence supports a scattering
interpretation.

To determine whether the exponential component is warranted, we also
fit a Gaussian to the same frequency-integrated profile and compare
the two models using the Bayesian information criterion (BIC). The BIC
balances fit quality against model complexity, so that the improvement
provided by the EMG must justify its additional tail parameter; the
model with the lower BIC is preferred. Both models are fitted to the bin-integrated profile by nonlinear
least squares. The optimisation settings, parameter bounds,
covariance estimates, and BIC definition are given in
\ref{app:fitting}. We define

\begin{equation}
\Delta\mathrm{BIC}
\equiv
\mathrm{BIC}_{\rm Gauss}
-
\mathrm{BIC}_{\rm EMG},
\label{eq:delta_bic}
\end{equation}
so that positive values favour the EMG. We adopt
$\Delta\mathrm{BIC}>6$ as a conservative threshold for retaining the
additional tail component. This corresponds approximately to an evidence ratio
of $20{:}1$ in favour of the EMG over the competing Gaussian model and is conventionally regarded as strong support \citep{kass1995}. Otherwise, the simpler Gaussian is retained.

Both models are integrated over the finite time bins before
comparison, rather than evaluated only at their centres, because the
fitted Gaussian width is comparable to the native 0.983-ms sampling.
For FRB~20210415C, the comparison gives
$\Delta\mathrm{BIC}=449$, overwhelmingly favouring the EMG. The
best-fitting parameters are
\begin{equation}
\sigint=0.93\pm0.05~\mathrm{ms},
\qquad
\tauemp=12.2\pm0.2~\mathrm{ms}.
\label{eq:fullband}
\end{equation}
These parameters characterise the frequency-integrated profile. In
particular, $\tauemp$ is a band-integrated empirical timescale and
therefore does not represent the scattering time at any single
observing frequency.
Although the EMG is strongly preferred over a Gaussian, the
full-band fit has

\begin{equation}
\chi^2_{\rm red,prof}
\equiv
\frac{\chi^2}{N_{\rm bin}-N_{\rm par}}
=
7.3,
\label{eq:profile_chi2}
\end{equation}
where
$\chi^2=\sum_k[(I_k-I_{{\rm mod},k})/\sigma_k]^2$,
$I_k$ and $I_{{\rm mod},k}$ are the observed and modelled intensities
in time bin $k$, and $\sigma_k$ is the adopted uncertainty in that
bin. Here, $N_{\rm bin}$ is the number of fitted time bins,
$N_{\rm par}$ is the number of free model parameters, and their
difference is the number of degrees of freedom. The elevated reduced
chi-squared indicates that the preferred EMG does not capture all
structure in the full-band profile.

Assuming that the excess residual variance can be represented by a
common scale factor, we multiply the formal covariance matrix by
$\chi^2_{\rm red,prof}=7.3$. Equivalently, the formal parameter
uncertainties are enlarged by
$\sqrt{\chi^2_{\rm red,prof}}=2.70$. The uncertainties quoted in
Equation~\ref{eq:fullband} already include this rescaling, while the
best-fitting parameter values are unchanged. The full-band fit is
shown in Figure~\ref{fig:waterfall}(a).

\FloatBarrier

\subsection{Frequency-resolved tail measurement}
\label{sec:frequency_analysis}

To resolve the tail across frequency, we analyse the interval
400.2--765.6~MHz. This interval extends from the lower edge of the
CHIME observing band to the frequency at which the Catalog~2 spectral
model has fallen to 4.8 per cent of its peak. We divide it into nine
contiguous sub-bands containing equal numbers of unmasked channels;
because the masked fraction varies with frequency, the sub-bands need
not have equal frequency widths. The low-frequency sub-bands provide
the strongest leverage on $\alpha$ because their tails are longest.
As these tails approach the end of the recorded time window, we test
the effect of possible truncation in \ref{app:injections}. The
sub-band boundaries are indicated by horizontal dashed lines in Figure~\ref{fig:waterfall}(b).

Each sub-band $j$ is assigned the flux-weighted effective frequency

\begin{equation}
\nu_{{\rm eff},j}
=
\frac{\sum_{i\in j} S_i\nu_i}
     {\sum_{i\in j} S_i},
\label{eq:effective_frequency}
\end{equation}
where the sums include only unmasked channels and $S_i$ is the
Catalog~2 running-power-law spectral model evaluated at channel
frequency $\nu_i$ \citep{chime2026}. The same
$\nu_{{\rm eff},j}$ is used to centre the drift correction and to
assign a representative frequency to the recovered sub-band tail
timescale. The time-integrated spectrum and spectral model $S(\nu)$
are shown in Figure~\ref{fig:waterfall}(c). Because brighter channels
contribute more strongly to the summed temporal profile, the
flux-weighted frequency is more representative of that profile than
the arithmetic mean. This convention is tested in
\ref{app:injections}.

Before summing the channels within each sub-band, we correct for any
residual time--frequency drift. Without this correction, channel
summation could artificially broaden the profile, with the additional
width absorbed into either $\sigint$ or $\tauemp$. Using the
per-channel arrival-time method of \citet{chamma2025}, we obtain 53
valid arrival-time measurements. These are overlaid as black points on
the dynamic spectrum in Figure~\ref{fig:waterfall}(b) and shown as a
function of frequency in Figure~\ref{fig:waterfall}(d). A linear fit
gives

\begin{equation}
\dtdnu
=
(-4.6\pm2.6)\times10^{-4}
~\mathrm{ms\,MHz^{-1}}.
\label{eq:burst_slope}
\end{equation}

The fitted slope differs from zero by only $1.8\sigma$, so no
significant drift is detected. We nevertheless apply the fitted
correction uniformly to prevent any residual drift from contributing
to the fitted Gaussian width or tail timescale. Channel $i$ in
sub-band $j$ is shifted by

\begin{equation}
\delta t_{ij}
=
-\left(\dtdnu\right)
\left(\nu_i-\nu_{{\rm eff},j}\right),
\label{eq:dedrift}
\end{equation}
where $\delta t_{ij}$ is the applied time shift and $\nu_i$ is the
channel frequency. The slope is measured from the original dynamic
spectrum, while the correction is applied only when constructing the
sub-band profiles. All subsequent analyses use the same dedrifted
profiles.

After dedrifting, we sum the unmasked channel time series within each
sub-band at every time sample to form a single sub-band-integrated
temporal profile. Fitting this profile with the bin-integrated EMG
yields an empirical tail timescale $\tau_{{\rm emp},j}$ and a
Gaussian-component width $\sigma_{{\rm int},j}$. These are effective
parameters of the finite sub-band profile rather than measurements
from an individual frequency channel.

A sub-band measurement is retained only if the EMG fit converges to a
finite, positive tail timescale, neither $\sigma_{{\rm int},j}$ nor
$\tau_{{\rm emp},j}$ reaches its numerical lower bound, and
$\chi^2_{\rm red,prof}\leq5$. The chi-squared condition is used only
as a quality-control threshold for the sub-band measurements; it is not
used to retain or reject the full-band EMG in the BIC
comparison.\footnote{The optimiser imposes
$\sigint,\tauemp\geq0.49$~ms, half the native sampling interval, to
prevent either parameter from collapsing toward zero. None of the
retained fits returns either parameter at this limit, so the reported
measurements are not determined by the numerical bound. Across the
nine sub-band fits, the narrowest recovered Gaussian-component width
is $\sigint=0.89$~ms, corresponding to a Gaussian FWHM of approximately
2.10~ms.}

The highest-frequency sub-band has only 43 per cent of its native
channels unmasked. We therefore exclude it from the primary
tail--frequency fit, leaving eight retained profiles. This
conservative choice is based on the limited channel coverage rather
than on failure to meet an EMG-fit quality criterion. The effect of including
this sub-band is quantified in \ref{app:selection}.

We associate each fitted tail timescale with the representative
frequency $\nu_{{\rm eff},j}$ defined above. The resulting pairs
$(\nu_{{\rm eff},j},\tau_{{\rm emp},j})$ are fitted with

\begin{equation}
\tauemp(\nu)
=
\tau_{\rm ref}
\left(
\frac{\nu}{\nu_{\rm ref}}
\right)^\alpha,
\label{eq:tail_law}
\end{equation}
evaluated at $\nu=\nu_{{\rm eff},j}$. Taking logarithms gives
$\ln\tau_{{\rm emp},j}
=\ln\tau_{\rm ref}
+\alpha\ln(\nu_{{\rm eff},j}/\nu_{\rm ref})$,
which we fit by weighted least squares. Here, $\tau_{\rm ref}$ is the
normalisation of the empirical tail--frequency relation and $\alpha$ is fitted freely. We
choose $\nu_{\rm ref}=\reffreq$, the midpoint of the
400.2--765.6~MHz interval used to construct the sub-bands. This choice
sets the value of $\tau_{\rm ref}$ but does not affect the fitted
index $\alpha$.

Because the power law is fitted to $\ln\tau_{{\rm emp},j}$, each
profile-fit uncertainty must be converted to an uncertainty in
logarithmic space. To first order in the fractional uncertainty,

\begin{equation}
\sigma_{\ln\tau,j}
\simeq
\frac{\sigma_{\tau,j}}{\tau_{{\rm emp},j}},
\end{equation}
where $\sigma_{\tau,j}$ is the profile-fit uncertainty on
$\tau_{{\rm emp},j}$. Each sub-band uncertainty includes the
profile-level residual-variance rescaling by that fit's own
$\chi^2_{\rm red,prof}$, following the procedure described in
Section~\ref{sec:data}. The corresponding regression weight is
$\omega_j=1/\sigma_{\ln\tau,j}^{2}$, so more precise sub-band
measurements contribute more strongly.

Because a broader Gaussian component can be partly compensated by a
shorter exponential tail, $\sigma_{{\rm int},j}$ and
$\tau_{{\rm emp},j}$ are partially covariant in an EMG fit. We
therefore perform two complementary analyses. In the free-width
analysis, both parameters vary independently in every sub-band. In the
fixed-width analysis, $\sigma_{{\rm int},j}$ is set to 1.12~ms, the
median of the eight retained free-width measurements, while
$\tau_{{\rm emp},j}$ remains free. The fixed-width analysis is a
sensitivity test rather than an independent measurement because the
common width is derived from the same data.

At the power-law regression stage, if the eight tail measurements show
more scatter about the fitted relation than represented by their
profile-fit uncertainties, we enlarge the formal uncertainties on
$\tau_{\rm ref}$ and $\alpha$ by
$\sqrt{\chi^2_{\rm red,PL}}$, leaving their best-fitting values
unchanged. This regression-level correction is separate from the
profile-level rescaling described above. The definition of this correction and a joint-profile
check of the width--tail covariance are given in
\ref{app:covariance}.

We also examine the frequency dependence of the Gaussian-component
widths returned by the free-width fits. We fit

\begin{equation}
\sigint(\nu)
=
\sigma_{\rm ref}
\left(
\frac{\nu}{\nu_{\rm ref}}
\right)^\beta,
\label{eq:width_law}
\end{equation}
where $\sigma_{\rm ref}$ is the fitted Gaussian-component width at
$\nu_{\rm ref}$ and $\beta$ is the width--frequency index. As
secondary descriptive diagnostics, we fit all nine width measurements
and, separately, the seven retained measurements with
$\nu_{{\rm eff},j}>440$~MHz; the latter fit isolates the
higher-frequency trend. Unlike the tail--frequency fits, these width
fits are unweighted. The reason for this choice, together with the sensitivity to the
adopted frequency limits and sub-band construction, is examined in
\ref{app:width_tests}; dependence on the assumed pulse-broadening
function is tested in \ref{app:pbf_tests}.

Throughout this subsection, $\tau_{{\rm emp},j}$ and
$\tau_{\rm ref}$ are empirical fit quantities. After the measured
frequency dependence and validation tests support a scattering
interpretation, we identify $\tauemp(\nu)$ with $\tausc(\nu)$ and
$\tau_{\rm ref}$ with $\tausc(\nu_{\rm ref})$.

\subsection{Scintillation search}
\label{sec:scint_method}

To search for Galactic diffractive scintillation, we form an on-pulse
spectrum by summing the Catalog~2 emission window after subtracting
the off-pulse mean from each frequency channel. We analyse four
contiguous runs of at least 300 unmasked channels spanning
600.0--610.7, 641.6--660.7, 664.3--693.2, and
700.7--714.4~MHz.

For each block, we compute the one-sided autocorrelation function
(ACF) of the fractional spectral deviations to a maximum frequency
lag of 2~MHz, following \citet{masui2015}. Zero lag is excluded from
the fit because it contains the radiometer-noise contribution. We
model the narrow ACF component as
\begin{equation}
{\rm ACF}_{\rm scint}(\delta\nu)
=
\frac{m^2}
{1+(\delta\nu/\Delta\nu_{\rm d})^2},
\end{equation}
where $\delta\nu$ is the frequency lag, $m$ is the modulation index,
and $\Delta\nu_{\rm d}$ is the decorrelation bandwidth. Before fitting
the narrow ACF component, we remove the broad burst spectral envelope
using the Gaussian-smoothing procedure described in
\ref{app:scint}. The exact ACF estimator, envelope treatment, and
uncertainty prescription are also given in \ref{app:scint}.

We classify a block as a scintillation detection only if both $m$ and
$\Delta\nu_{\rm d}$ exceed their uncertainties by $3\sigma$ and
$\Delta\nu_{\rm d}$ exceeds 1.5 native channels. We calibrate the
sensitivity and frequency resolution of this criterion through
injection--recovery tests performed on the observed dynamic spectrum;
the injection procedure is described in \ref{app:scint}.

\section{Results}
\label{sec:results}

\subsection{Frequency-resolved scattering law}
\label{sec:scattering_result}

\begin{figure*}[!t]
\centering
\includegraphics[width=\textwidth]
{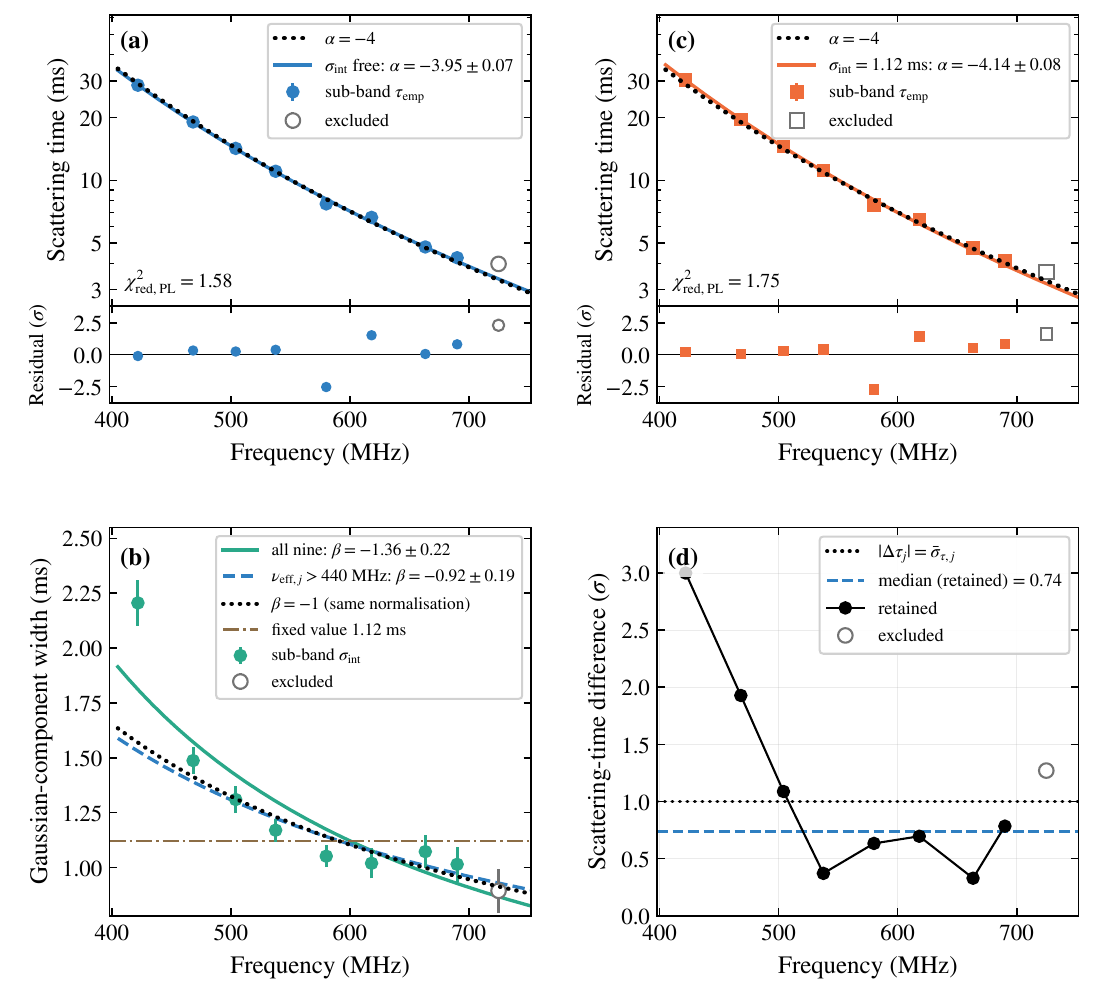}
\caption{
Frequency-resolved tail measurements and Gaussian-width diagnostics
for FRB~20210415C.
\emph{(a)} Empirical sub-band tail timescales
$\tau_{{\rm emp},j}$ obtained when both
$\sigma_{{\rm int},j}$ and $\tau_{{\rm emp},j}$ are free in every
EMG fit. The solid curve is the fitted power law, and the dotted curve
is the reference $\alpha=-4$ relation normalised to the fitted
$\tau_{\rm ref}$ at $\nu_{\rm ref}$. The lower strip shows the
normalised residual
$(\tau_{{\rm emp},j}-\tau_{{\rm fit},j})/\sigma_{\tau,j}$, where
$\tau_{{\rm fit},j}$ is the power-law prediction at
$\nu_{{\rm eff},j}$.
\emph{(b)} Gaussian-component widths
$\sigma_{{\rm int},j}$ obtained from the free-width EMG fits used in
panel~(a). The solid line is the fit to all nine sub-bands,
$\beta=-1.36\pm0.22$, and the dashed line is the fit to the retained
sub-bands with $\nu_{{\rm eff},j}>440$~MHz,
$\beta=-0.92\pm0.19$. The dotted line shows a reference
$\beta=-1$ relation with the same normalisation as the dashed fit, and
the dash-dotted line marks the common width
$\sigma_{\rm int}=1.12$~ms imposed in panel~(c).
\emph{(c)} Tail timescales recovered after refitting the same sub-band
profiles with $\sigma_{\rm int}=1.12$~ms fixed and
$\tau_{{\rm emp},j}$ free. The curves and lower residual strip are
defined as in panel~(a).
\emph{(d)} Change in the recovered tail timescale between the
free-width and fixed-width fits,
$|\Delta\tau_j|/\bar{\sigma}_{\tau,j}$, where
$\Delta\tau_j\equiv
\tau_{{\rm emp},j}^{\rm free}
-\tau_{{\rm emp},j}^{\rm fixed}$ and
$\bar{\sigma}_{\tau,j}\equiv
(\sigma_{\tau,j}^{\rm free}+\sigma_{\tau,j}^{\rm fixed})/2$.
The dotted horizontal line marks a change equal to one mean
profile-fit uncertainty, and the dashed line marks the median value
of 0.74 for the retained sub-bands. Open grey symbols show the highest-frequency sub-band for completeness.
It is excluded from the tail--frequency fits in panels~(a) and (c) and
from the retained-sub-band median in panel~(d), but is included in the
all-nine-sub-band width fit in panel~(b). In panels~(a) and (c),
$\chi^2_{\rm red,PL}$ refers to the two-parameter power-law fit to
eight retained tail measurements, giving six degrees of freedom, and
not to the individual EMG profile fits.
}
\label{fig:subband}
\end{figure*}

The fitted empirical tail timescale increases systematically toward
lower frequencies in both the free-width analysis
(Figure~\ref{fig:subband}(a)) and the fixed-width sensitivity test
(Figure~\ref{fig:subband}(c)). The corresponding power-law fits give

\begin{equation}
\begin{aligned}
(\alpha,\tau_{\rm ref})_{\rm free}
&=
(-3.95\pm0.07,\,7.98\pm0.11~\mathrm{ms}),\\
(\alpha,\tau_{\rm ref})_{\rm fixed}
&=
(-4.14\pm0.08,\,7.90\pm0.12~\mathrm{ms}),
\end{aligned}
\label{eq:scattering_results}
\end{equation}
at the reference frequency $\nu_{\rm ref}=\reffreq$. The corresponding
fits have $\chi^2_{\rm red,PL}=1.58$ and $1.75$, respectively. Here,
$\chi^2_{\rm red,PL}$ is the reduced chi-squared of the weighted
power-law fit to eight retained sub-band measurements with two free
parameters, giving six degrees of freedom; its definition and the
associated uncertainty rescaling are given in
\ref{app:covariance}. Both indices are consistent with the commonly
adopted $\nu^{-4}$ scattering law.

The Gaussian-component widths returned by the free-width EMG fits
also decrease with frequency, as shown in
Figure~\ref{fig:subband}(b). Restricting the fit to the seven retained
sub-bands with $\nu_{{\rm eff},j}>440$~MHz gives
$\beta=-0.92\pm0.19$, corresponding to an approximately
$\nu^{-1}$ trend. The remaining retained sub-band, at
$\nu_{{\rm eff}, j}\simeq422$~MHz, has
$\sigint=2.21\pm0.10$~ms and lies $4.8\sigma$ above the extrapolation
of this relation. For comparison, fitting all nine available width measurements gives
the steeper index $\beta=-1.36\pm0.22$. The difference between the two
fitted indices is therefore driven primarily by the low-frequency
excess, and we do not interpret the all-nine fit as a unique
band-wide width law.

The alternative frequency ranges, sub-band configurations, and
injection tests presented in \ref{app:width_tests} show that the
low-frequency excess persists and is not produced by the finite
recording window or by the width-recovery procedure within the adopted
EMG model. We therefore treat the approximately $\nu^{-1}$ behaviour
above 440~MHz as a secondary empirical profile-shape trend. Its
dependence on the assumed pulse-broadening function is examined in
\ref{app:pbf_tests}, and its possible physical interpretation is
discussed in Section~\ref{sec:discussion}. Table~\ref{tab:summary} lists the principal burst and fit quantities.

\begin{table}[!t]
\centering
\caption{
Summary of the full-band and frequency-resolved fit results for
FRB~20210415C. The sub-band power-law parameters are derived from the
eight retained dedrifted profiles and are quoted at
$\nu_{\rm ref}=\reffreq$. The fixed-width sensitivity test uses
$\sigint=1.12$~ms in every sub-band. The full-band and sub-band
profile-fit uncertainties include the profile-level residual-variance
rescaling described in Section~\ref{sec:data}; the uncertainties on
$\alpha$ and $\tau_{\rm ref}$ also include the power-law-level
rescaling described in \ref{app:covariance}. The uncertainty on
$\beta$ is obtained from the unweighted log-space fit and is scaled by
the residual scatter.
}
\label{tab:summary}
\begin{tabular}{lc}
\hline
Quantity & Value \\
\hline
Full-band $\Delta\mathrm{BIC}$ & 449 \\
Full-band $\sigint$ (ms) & $0.93\pm0.05$ \\
Full-band $\tauemp$ (ms) & $12.2\pm0.2$ \\
$\alpha$ (free-width fit) & $-3.95\pm0.07$ \\
$\tau_{\rm ref}$ (free-width fit; ms) & $7.98\pm0.11$ \\
$\alpha$ (fixed-width fit) & $-4.14\pm0.08$ \\
$\tau_{\rm ref}$ (fixed-width fit; ms) & $7.90\pm0.12$ \\
$\beta$ (descriptive fit above 440~MHz) & $-0.92\pm0.19$ \\
\hline
\end{tabular}
\end{table}
The full-band value, $\tauemp=12.2\pm0.2$~ms, is fitted to the
frequency-integrated profile and is therefore an effective
band-integrated quantity, rather than the tail timescale at a single
frequency. The width treatment has only a modest effect on the sub-band
measurements. Across the retained sub-bands, fixing $\sigint$ changes
$\tau_{{\rm emp},j}$ by a median of 0.74 times the mean of the
corresponding free-width and fixed-width profile-fit uncertainties
(Figure~\ref{fig:subband}(d)). The largest change occurs in the
lowest-frequency sub-band, but the free-width and fixed-width
analyses recover consistent power-law indices. The largest drift
correction across a retained sub-band is
$2.4\times10^{-2}$~ms, more than two orders of magnitude below the
shortest fitted tail timescale. Thus, neither the treatment of the
Gaussian-component width nor residual drift smearing drives the
observed frequency dependence.

The validation tests in \ref{app:robustness} lead to the same
conclusion. The finite recording window changes the recovered index
by less than 0.02, while the flux-weighted sub-band frequencies reduce
the recovered-index offset by approximately 0.04 relative to
arithmetic-mean frequencies. Varying the DM by up to
$\pm0.25~\pccm$ changes $\alpha$ by at most 0.02 and
$\tau_{\rm ref}$ by at most 0.03~ms, while including the
highest-frequency sub-band gives $\alpha=-3.91\pm0.08$. A joint fit
to the eight retained profiles, allowing the Gaussian-component width
and centre to vary independently in each sub-band, gives
$\alpha=-3.95\pm0.06$ and
$\tau_{\rm ref}=7.97\pm0.09$~ms. As a negative control, synthetic
profiles with a frequency-independent asymmetric tail recover
$\alpha=0.01\pm0.04$, showing that the estimator does not itself
impose a scattering-like frequency dependence.

Taken together, the measured $\nu^{-4}$-like dependence and the
validation tests support identifying $\tauemp(\nu)$ with
$\tausc(\nu)$ within the adopted exponential pulse-broadening model.
At the analysis reference frequency,
$\nu_{\rm ref}=\reffreq$, the corresponding scattering times are
$7.98\pm0.11$~ms for the free-width fit,
$7.90\pm0.12$~ms for the fixed-width fit, and
$7.97\pm0.09$~ms for the joint-profile fit. To facilitate comparison
with measurements quoted at 600~MHz, we rescale the same fitted
relations without refitting the data. The resulting scattering times
are $7.12\pm0.11$~ms, $7.01\pm0.12$~ms, and
$7.11\pm0.09$~ms, respectively. The uncertainties on the rescaled
values include the covariance between $\tau_{\rm ref}$ and $\alpha$
(\ref{app:covariance}).

\subsection{Scintillation constraints}
\label{sec:scint_result}

Because the tail--frequency relation supports a scattering
interpretation but does not locate the dominant screen, we use the
scintillation search described in
Section~\ref{sec:scint_method} as a complementary diagnostic. The four
analysed channel blocks are centred at 605, 651, 679, and 708~MHz;
the detailed ACF construction and injection--recovery calibration are
given in \ref{app:scint}.

For temporal broadening and scintillation produced by the same
Galactic screen, the decorrelation bandwidth is related approximately
to the Milky Way pulse-broadening time by
$2\pi\Delta\nu_{\rm d}\tau_{\rm MW}=C_1$, where $C_1$ is a
dimensionless factor of order unity that depends on the scattering
geometry and fluctuation spectrum
\citep{masui2015,sammons2023}. Adopting $C_1=1$, the NE2001 and
YMW16 pulse-broadening estimates for this sightline
\citep{cordes2002,yao2017} correspond to
$\Delta\nu_{\rm d}\simeq44$ and $8$~kHz, respectively, at
\reffreq. Assuming $\Delta\nu_{\rm d}\propto\nu^4$, these become
51--96~kHz and 10--18~kHz across the four channel blocks. The
NE2001-predicted bandwidths therefore span approximately two to four
native 24.4-kHz channels, whereas the YMW16-predicted bandwidths lie
below one channel.

No block meets the scintillation-detection criterion, and no fitted
decorrelation bandwidth differs from zero by more than
$1.2\sigma$. Under the adopted injection model, NE2001-scale
scintillation would be recovered with the correct bandwidth in at
least 97 per cent of trials for $m\geq0.5$, but in only 66 and
11 per cent of trials for $m=0.4$ and $0.3$, respectively. The
non-detection therefore disfavours strong NE2001-scale modulation,
while weaker modulation remains allowed. YMW16-scale bandwidths
cannot be resolved at the native channel width; however, injections
with $m\geq0.7$ produce an unresolved, channel-limited signature in
at least 96 per cent of trials. Strong YMW16-scale modulation is
therefore also disfavoured, although its decorrelation bandwidth
cannot be measured.

The on-pulse spectra contain weak correlated structure, but no
associated decorrelation bandwidth is significantly measured. We
therefore cannot identify this structure as diffractive scintillation;
intrinsic or instrumental spectral structure remains possible. The
fitted ACF amplitudes, off-pulse controls, and injection tests are
reported in \ref{app:scint}.

\section{Discussion and conclusions}
\label{sec:discussion}

The principal result is a direct measurement of a
scattering-frequency index consistent with the commonly
adopted $\alpha=-4$ law. Unlike catalogue analyses that fix the
exponent, the present analysis derives it from the resolved
frequency evolution of the empirical tail.
The free-width, fixed-width, and joint-profile analyses give
consistent results, while the injection--recovery tests in
\ref{app:robustness} show that the steep frequency dependence is not
produced by width--tail covariance, truncation of the low-frequency
tails, or the sub-band frequency assignment.
A frequency-independent asymmetric control also recovers an
index consistent with zero.
These results support identifying $\tauemp(\nu)$ with a
scattering timescale $\tausc(\nu)$ within the adopted exponential
pulse-broadening model.
An index near $-4$ does not, however, uniquely determine the
scattering geometry or the spectrum of electron-density fluctuations.

The fitted Gaussian-component width also decreases with frequency.
Above 440~MHz, the measurements follow approximately
$\sigint\propto\nu^{-0.92\pm0.19}$,
qualitatively consistent with the inverse
duration--frequency scaling predicted by the triggered relativistic
dynamical model \citep{rajabi2020} and with temporal narrowing
discussed in triggered-superradiance models \citep{houde2019}. The lowest-frequency retained measurement,
$\sigint=2.21\pm0.10$~ms at approximately 422~MHz, lies
$4.8\sigma$ above the extrapolation of this trend. In the primary
configuration, this point drives the steeper all-band index and
suggests a low-frequency departure from the higher-frequency
relation. The tests in
\ref{app:width_tests} show that this low-frequency excess is not
generated by the finite recording window or by the width-recovery
procedure within the adopted EMG model.
Alternative-response injections in \ref{app:pbf_tests}
nevertheless show that a non-exponential pulse-broadening function,
such as that expected from an extended scattering medium, can produce
a similar increase: the EMG then absorbs part of the rounded
profile onset into $\sigint$ and can bias the inferred
$\alpha$ toward shallower values
\citep{williamson1972,bhat2004}.
Because $\sigint$ may also contain unresolved temporal
structure and instrumental broadening, we regard the approximately
$\nu^{-1}$ higher-frequency trend as a descriptive profile-shape
result rather than a unique measurement of the intrinsic emission
timescale or mechanism.

At \reffreq, the measured scattering time is
$\tausc\simeq8.0$~ms. Along this sightline, the NE2001 and YMW16
models predict Galactic pulse-broadening times of approximately
$3.6~\mu\mathrm{s}$ and $19~\mu\mathrm{s}$, respectively
\citep{cordes2002,yao2017}. The YMW16 estimate is based on the
empirical pulsar $\tau$--DM relation of
\citet{krishnakumar2015}. The measured value exceeds these predictions
by factors of approximately $2\times10^3$ and $4\times10^2$,
strongly disfavouring the Milky Way as the dominant source of
the broadening. The temporal broadening alone cannot distinguish
among a host-galaxy interstellar medium, a circumsource environment,
or an intervening scattering screen.

Scintillation provides a complementary diagnostic when
temporal broadening and Galactic spectral modulation arise from
distinct plasma screens \citep{masui2015,sammons2023}.
Under the adopted injection model, resolved scintillation at
the NE2001-predicted bandwidth with modulation index $m\geq0.5$ would
have been recovered in at least 97 per cent of trials. Its
non-detection therefore disfavours strong modulation on this scale,
although weaker NE2001-scale modulation with $m\lesssim0.4$ remains
allowed. The narrower YMW16-predicted bandwidths cannot be resolved at the
native channel width. Injections with $m\geq0.7$ nevertheless produce
an unresolved, channel-limited signature in at least 96 per cent of
trials, so strong modulation on this scale is disfavoured even though
the decorrelation bandwidth itself cannot be measured. The absence of a resolved feature could reflect weak modulation at
the NE2001-predicted bandwidth or partial quenching by angular
broadening from a second, extragalactic screen. The weak 120--230-kHz structure in the
on-pulse spectrum cannot be identified as scintillation because its
decorrelation bandwidth is poorly constrained. The scintillation
search therefore neither determines the screen distance nor
establishes a particular two-screen geometry.

This analysis concerns a single exceptionally bright burst and
therefore cannot establish whether the scattering timescale and its
frequency dependence are stable between bursts from
FRB~20210219C. Its main significance is that the frequency dependence
of the temporal tail is measured rather than imposed. Applying the
same approach to additional bright, broadband bursts would allow
catalogue analyses to test the scattering law instead of imposing
$\alpha=-4$ and, for repeating sources, determine whether the
scattering properties vary between bursts. Future observations with
baseband resolution and secure host localisation would further
clarify the Gaussian-width evolution and help locate the dominant
scattering screen.

\FloatBarrier
\appendix
\setcounter{figure}{0}

\section{Profile fitting and model selection}
\label{app:fitting}

This appendix gives the numerical details of the Gaussian and EMG
profile fits and the BIC model comparison described in
Section~\ref{sec:data}.

The independent full-band and sub-band Gaussian and EMG profile fits
are performed by nonlinear least squares using
\texttt{scipy.optimize.curve\_fit}. Because finite parameter bounds
are imposed, \texttt{curve\_fit} selects the trust-region-reflective
algorithm (\texttt{method=`trf'}). The analysis was performed with
SciPy~1.18 and NumPy~2.5. For each profile, the fit minimises
$\chi^2=\sum_{n=1}^{N}
[(I_n-I_{{\rm mod},n})/\sigma_{\rm off}]^2$
over all $N=162$ time bins, where $\sigma_{\rm off}$ is the standard
deviation of the 74 leading off-pulse samples, which end five samples
before the Catalog~2 emission window. This profile-specific
uncertainty is applied to every time bin and treated as an absolute
uncertainty. Each model is numerically averaged over the 0.983-ms time
bins using 32 equally spaced samples per bin.

The full-band profile is the sum of all unmasked channels over
400.2--799.8~MHz and is fitted without dedrifting. The sub-band
profiles are the dedrifted sums constructed in
Section~\ref{sec:frequency_analysis}. The full-band and sub-band fits
use the same optimiser, bin averaging, noise model, parameter bounds,
and covariance prescription; only their starting values differ.

The EMG has five free parameters,
$(A,\mu,\sigint,\tauemp,C)$, with
$A\geq0$, $0\leq\mu\leq158.3$~ms,
$0.49\leq\sigint\leq30$~ms, and
$0.49\leq\tauemp\leq500$~ms, while $C$ is unbounded.
The Gaussian has four free parameters,
$(A,\mu,\sigint,C)$, with the same bounds except that
$\sigint\leq100$~ms.

For the full-band fits, the initial baseline $C_0$ is the median of
the leading off-pulse samples, and the initial amplitude is
$A_0=\Delta t\sum_{n=1}^{N}(I_n-C_0)$. The initial centre is the
profile peak for the Gaussian and 1~ms before the peak for the EMG.
The Gaussian fit starts from $\sigint=3$~ms, while the EMG fit starts
from $\sigint=1$~ms and $\tauemp=5$~ms.

The free-width sub-band fits start from
$A_0=\Delta t\sum_{n=1}^{N}I_n$, $\mu_0=79$~ms,
$\sigma_{{\rm int},0}=1$~ms, $\tau_{{\rm emp},0}=10$~ms, and $C_0=0$.
The fixed-width fits start from the corresponding free-width
solutions, with $\sigint$ held at 1.12~ms. A fit is treated as failed
if the optimiser does not converge within $2\times10^5$ model
evaluations under its default tolerances.

The reported results use these default starting values. To test their
sensitivity to initialisation, we repeated every EMG fit using all 60
combinations of
$\sigint\in\{0.6,1,2,4\}$~ms,
$\tauemp\in\{1,5,10,20,40\}$~ms, and
$\mu\in\{75,79,83\}$~ms. For the fixed-width fits, we used the 15
combinations of $\tauemp$ and $\mu$. Every restart returned the same
minimum: $\chi^2$ never decreased by more than $3\times10^{-7}$, and
no fitted parameter changed by more than $2\times10^{-3}$ of its
quoted uncertainty.

The formal parameter covariance is the Gauss--Newton estimate
$(J^{\mathsf T}J)^{-1}$, where $J$ is the Jacobian of the residuals
normalised by the absolute uncertainty $\sigma_{\rm off}$. Following
Section~\ref{sec:data}, each covariance matrix is multiplied once by
that profile's $\chi^2_{\rm red,prof}$. This leaves the best-fitting
parameters unchanged and enlarges their standard uncertainties by
$\sqrt{\chi^2_{\rm red,prof}}$.

For the Gaussian--EMG model comparison, the BIC is evaluated using
the Gaussian-likelihood form with the common noise variance estimated
from the residuals,

\begin{equation}
\mathrm{BIC}
=
N\ln\left(\frac{\mathrm{RSS}}{N}\right)
+
k\ln N,
\qquad
\mathrm{RSS}
=
\sum_{n=1}^{N}
\left(I_n-I_{{\rm mod},n}\right)^2,
\label{eq:bic_definition}
\end{equation}
where $N=162$ is the number of fitted time bins and $k=5$ and $4$ for
the EMG and Gaussian, respectively. This form uses neither
$\sigma_{\rm off}$ nor the subsequent
$\chi^2_{\rm red,prof}$ covariance rescaling. Because
$\sigma_{\rm off}$ is constant within a profile, minimising $\chi^2$
and minimising the RSS give the same best-fitting parameters.

For comparison, the fixed-variance form
$\mathrm{BIC}=\chi^2+k\ln N$ gives
$\Delta\mathrm{BIC}\simeq1.8\times10^4$ when the original off-pulse
uncertainty is used and
$\Delta\mathrm{BIC}\simeq2.4\times10^3$ after the profile-level
residual-variance rescaling. The strong preference for the EMG is
therefore insensitive to the adopted BIC convention.

\section{Robustness of the scattering-law measurement}
\label{app:robustness}

This appendix tests the principal analysis choices that could affect
the recovered tail--frequency relation. We examine the finite
recording window and effective-frequency convention, the adopted
dispersion measure and sub-band selection, the power-law uncertainty
treatment and width--tail covariance, and the recovery and model
dependence of the Gaussian-component width.

\subsection{Recording-window truncation and effective-frequency tests}
\label{app:injections}

The tail is longest at the bottom of the observing band and is not
fully contained within the recorded time window. The
lowest-frequency profile contains approximately 82~ms after the tail
onset, corresponding to about 2.3 e-folding times of the fitted
scattering law at the bottom of the band.
If fitting only a partial tail biases $\tauemp$ downward, the
low-frequency measurements could be underestimated and the recovered
frequency dependence artificially flattened.

We test this possibility by injecting synthetic bursts with known
$(\tau_{\rm ref},\alpha)$ into the observed time--frequency grid. The
injections use the measured spectral envelope, channel mask, sampling,
and sub-band construction and are processed with the same dedrifting
and EMG-fitting procedure as the data. We compare recovery from the
observed 162-sample window with recovery from an extended window that
contains the full injected tail.

\begin{figure*}[!t]
\centering
\includegraphics[width=\textwidth]
{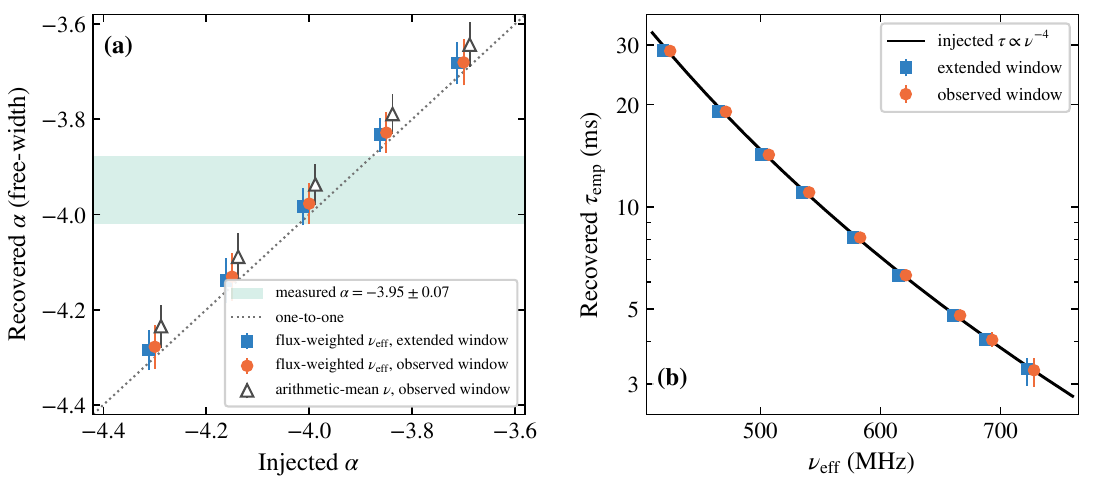}
\caption{
Injection--recovery tests of the scattering-frequency index.
\emph{(a)} Recovered versus injected $\alpha$ for the free-width
analysis. Points and error bars show the mean and standard deviation
of 100 trials at each injected value. Blue squares and orange circles
use the flux-weighted sub-band frequencies in the extended and
observed windows, respectively; open triangles use arithmetic-mean
frequencies in the observed window. The dotted line marks one-to-one
recovery, and the shaded band shows the measured free-width value,
$\alpha=-3.95\pm0.07$. The observed- and extended-window recoveries
agree to within 0.02, while the arithmetic-mean convention shifts the
recovered index by approximately 0.04 relative to the flux-weighted
convention.
\emph{(b)} Recovered sub-band tail timescales for injections with
$\alpha=-4$. The black curve is the injected relation; blue squares
and orange circles show the median recovery from the extended and
observed windows, respectively, with error bars spanning the
16th--84th percentiles of 100 trials. Their agreement shows that
truncation of the low-frequency tails does not alter the recovered
frequency dependence.
}
\label{fig:alpha_inject}
\end{figure*}

Over $-4.3\leq\alpha\leq-3.7$, the indices recovered from the two
windows differ by less than 0.02
(Figure~\ref{fig:alpha_inject}(a)). The recovered sub-band
timescales also follow the injected $\nu^{-4}$ relation despite the
truncated low-frequency profiles
(Figure~\ref{fig:alpha_inject}(b)). The finite recording window
therefore has a negligible effect on the measured index.

The same injections test the representative frequency assigned to
each finite sub-band. For an injected $\alpha=-4$, arithmetic-mean
sub-band frequencies recover approximately $-3.94$, whereas the
flux-weighted frequencies of
Equation~\ref{eq:effective_frequency} recover approximately $-3.98$.
Replacing the Catalog~2 spectral model with a Gaussian fitted to the
time-integrated spectrum changes the recovered index by less than
0.01. The adopted frequency convention therefore reduces the
estimator offset to well below the uncertainty on the measured
$\alpha$.

\subsection{Sensitivity to dispersion measure and sub-band selection}
\label{app:selection}

We repeat the full analysis over
$\mathrm{DM}=526.05$--$526.55~\pccm$, using offsets of
$0$, $\pm0.05$, $\pm0.10$, and $\pm0.25~\pccm$ from the adopted
value. At each trial DM, the data are rededispersed, the drift slope is
remeasured, and the free-width and fixed-width analyses are repeated.
The largest offset brackets the $0.21~\pccm$ difference between the
adopted DM and the real-time detection DM of $526.09~\pccm$
\citep{chime2026}. Across the tested range,
$\alpha$ changes by at most 0.02 and $\tau_{\rm ref}$ by at most
0.03~ms.

Including the highest-frequency sub-band gives
$\alpha=-3.91\pm0.08$. Leave-one-sub-band-out fits span
$-3.99\leq\alpha\leq-3.88$ when starting from the eight retained
profiles and
$-3.95\leq\alpha\leq-3.84$ when starting from all nine profiles.

As an alternative bandwidth choice, we raise the lower limit to
428.6~MHz and divide 428.6--765.6~MHz into eight sub-bands. This gives
$\alpha=-3.97\pm0.12$ for the free-width analysis and
$\alpha=-4.11\pm0.13$ for the fixed-width analysis. None of the
tested DM, bandwidth, or sub-band-selection choices changes the
scattering interpretation.

\subsection{Power-law uncertainties, joint-profile fit, and negative
control}
\label{app:covariance}

We first define the uncertainty treatment used for the two-step
power-law regression, then test the width--tail covariance with a
joint fit to the eight retained profiles, and finally apply the
estimator to a frequency-independent asymmetric control.

For each treatment of the Gaussian-component width, the eight pairs
$(\nu_{{\rm eff},j},\tau_{{\rm emp},j})$ are fitted with
Equation~\ref{eq:tail_law}. We define
\begin{equation}
\chi^2_{\rm red,PL}
=
\frac{\chi^2_{\rm PL}}{N-p},
\label{eq:powerlaw_chi2}
\end{equation}
where $\chi^2_{\rm PL}$ is the weighted sum of squared residuals in
logarithmic space, $N=8$ is the number of retained measurements, and
$p=2$ corresponds to the fitted parameters
$(\tau_{\rm ref},\alpha)$. Each power-law fit therefore has six
degrees of freedom.

When $\chi^2_{\rm red,PL}>1$, the formal covariance matrix of
$(\tau_{\rm ref},\alpha)$ is multiplied by
$\chi^2_{\rm red,PL}$, equivalent to enlarging the standard
uncertainties by $\sqrt{\chi^2_{\rm red,PL}}$. This rescaling accounts
for scatter about the power law that exceeds that represented by the
individual profile-fit uncertainties and leaves the best-fitting
values unchanged.

The two-step analysis first fits each sub-band profile independently
and then fits the recovered tail timescales with
Equation~\ref{eq:tail_law}. Because
$\sigma_{{\rm int},j}$ and $\tau_{{\rm emp},j}$ are partially
covariant within an EMG fit, we also fit all eight retained temporal
profiles simultaneously. Their tail timescales share
$(\tau_{\rm ref},\alpha)$ through Equation~\ref{eq:tail_law}, while
the Gaussian-component width, centre, amplitude, and baseline vary
independently between sub-bands.
Each sub-band has its own Gaussian centre because the fitted
centre of the lowest-frequency sub-band lies $0.53\pm0.08$~ms later
than the median of the other seven, consistent with the late
low-frequency arrival times in Figure~\ref{fig:waterfall}(d);
imposing a common centre would force this offset into the fitted
tail.

The likelihood is Gaussian,
$\ln\mathcal{L}
=-\tfrac{1}{2}\sum_j\sum_k
[(I_{jk}-I_{{\rm mod},jk})/s_j]^2$,
summed over the 162 bins of each retained profile. Here, $s_j$ is the
off-pulse uncertainty of sub-band $j$ multiplied by
$\sqrt{\chi^2_{\rm red,prof}}$ from its free-width fit, using the same
residual-variance rescaling as in the two-step analysis.
Each EMG model is averaged analytically over every time bin
using the difference of its cumulative distribution function at the
bin edges. This differs from the 32-point numerical average used in
\ref{app:fitting}; for these profiles, the two prescriptions differ
by less than $2\times10^{-5}$ of the profile peak.

The 34 parameters are $\ln\tau_{\rm ref}$ and $\alpha$, shared by all
sub-bands, together with an amplitude, Gaussian width, centre, and
baseline for each sub-band.
The priors on $\ln\tau_{\rm ref}$ and $\alpha$ are flat and
improper. The priors on the per-sub-band parameters are uniform within
the bounds of \ref{app:fitting}, and every predicted
$\tau_{{\rm emp},j}$ is required to be at least 0.49~ms.
We first maximise the likelihood with
\texttt{scipy.optimize.least\_squares}, starting from the two-step
free-width solution, and then sample the posterior with
\texttt{emcee} \citep{foremanmackey2013}. We use 96 walkers
initialised in a small ball around the maximum-likelihood solution,
with a mixture of differential-evolution (80 per cent) and
differential-evolution snooker (20 per cent) moves. We run 16\,000
steps, discard the first 2000 as burn-in, and thin by 10, retaining
134\,400 samples. The integrated autocorrelation time is at most 143
steps for any parameter (124 for $\ln\tau_{\rm ref}$ and $\alpha$), so
the retained chain spans at least 98 autocorrelation times, with an
effective sample size of about 9400.

Marginalising over the 32 sub-band parameters, the posterior medians
and 16th--84th percentile intervals are
\begin{equation}
\alpha=-3.95\pm0.06,
\qquad
\tau_{\rm ref}
=
7.97\pm0.09~\mathrm{ms}.
\label{eq:joint_result}
\end{equation}
The joint result agrees with the two-step measurement, showing that
the recovered frequency dependence is not driven by the width--tail
covariance.

As a negative control, we analyse 100 synthetic asymmetric
profiles with the observed spectral envelope, noise level, sampling,
and characteristic tail time, but with a frequency-independent tail.
The recovered indices have mean and standard deviation
$\alpha=0.01\pm0.04$. Otherwise equivalent
$\tauemp\propto\nu^{-4}$ injections give
$\alpha=-3.98\pm0.04$, also quoted as the mean and standard deviation
over 100 realisations.
The estimator therefore recovers the frequency dependence present in
the signal rather than assigning a scattering-like index to every
asymmetric profile.

\subsection{Sensitivity and recovery of the Gaussian-component width}
\label{app:width_tests}

The width--frequency relation is defined by
Equation~\ref{eq:width_law}. The fit to all nine width measurements
gives $\beta=-1.36\pm0.22$, whereas the seven retained measurements
with $\nu_{{\rm eff},j}>440$~MHz give
$\beta=-0.92\pm0.19$. The highest-frequency profile is included in
the all-nine width fit because its exclusion from the tail-frequency
analysis is based on its limited valid-channel fraction rather than a
failure of its Gaussian-width fit.

The lowest-frequency retained measurement,
$\sigint=2.21\pm0.10$~ms at
$\nu_{{\rm eff},j}\simeq422$~MHz, lies $4.8\sigma$ above the
extrapolated higher-frequency relation. Here, the denominator is the
quadrature sum of the measured-width uncertainty and the uncertainty
of the extrapolated relation, including the covariance between
$\sigma_{\rm ref}$ and $\beta$.
The all-nine index is therefore strongly influenced by the
lowest-frequency measurement and should not be interpreted as a
unique band-wide width law.

We assess the sensitivity of this result by repeating the free-width
analysis over 770 alternative frequency ranges and sub-band
constructions. These comprise every combination of a lower frequency
limit of 400.2~MHz or 405--450~MHz in 5-MHz steps (11 values), an upper
frequency limit of 720--760~MHz in 5-MHz steps or 765.6~MHz
(10 values), and 6--12 sub-bands (7 values), giving
$11\times10\times7=770$ configurations. The primary configuration is
included in this grid. For each configuration, the sub-bands are
rebuilt with equal numbers of unmasked channels, and the profiles are
dedrifted and fitted with the same bin-integrated EMG. The recovered
Gaussian-component widths are then fitted with
Equation~\ref{eq:width_law}.

The median recovered index is $\beta=-1.06$, with a median formal
$1\sigma$ uncertainty of 0.21 and a 16th--84th percentile
configuration range of $-1.26\leq\beta\leq-0.96$. Because the
configurations reuse the same data, this interval measures sensitivity
to the analysis choices rather than an independent statistical
uncertainty. Configurations including more data below 440~MHz
generally give steeper indices (median $\beta=-1.24$ for the 276
configurations with at least one sub-band at
$\nu_{{\rm eff},j}<440$~MHz, compared with $-1.01$ for the remaining
494), while the higher-frequency trend remains close to $\nu^{-1}$.

All width--frequency fits in this paper are unweighted least-squares
fits in logarithmic space, with parameter uncertainties scaled by the
residual scatter.
We do not weight by the profile-fit uncertainties on
$\sigma_{{\rm int},j}$ because the scatter about a single power law
substantially exceeds that represented by those uncertainties. A fit
weighted by these uncertainties has
$\chi^2_{\rm red}=4.9$ for all nine measurements and
$\chi^2_{\rm red}=1.5$ for the seven measurements above 440~MHz.
Weighting by these uncertainties, with the same
$\sqrt{\chi^2_{\rm red}}$ rescaling used for the tail fits, gives
$\beta=-1.07\pm0.20$ above 440~MHz and places the lowest-frequency
measurement $4.6\sigma$ above the extrapolation.
This significance uses the same quadrature combination of the
measured-width and extrapolation uncertainties as the unweighted
comparison.
The weighted and unweighted results are therefore consistent.

We test the width recovery using synthetic bursts with the observed
spectral envelope, channel mask, sampling, noise level, and sub-band
construction. The injected Gaussian-component width is
$\sigint(\nu)=
\sigma_0(\nu/\nu_{\rm ref})^{\beta_{\rm inj}}$, with
$\sigma_0=1.12$~ms, the common width of the fixed-width analysis, at
$\nu_{\rm ref}=\reffreq$ and
$\beta_{\rm inj}=0$ (frequency independent) or $-1$, while
$\tauemp=\tau_{\rm ref}(\nu/\nu_{\rm ref})^{-4}$ with
$\tau_{\rm ref}=7.98$~ms. We repeat the analysis using lower frequency
limits of 400.2, 420, and 440~MHz and 6, 9, and 12 sub-bands, with 100
noise realisations for each width law and configuration.

The injected indices are recovered to within approximately 0.04: the
configuration means lie in
$-0.04\leq\beta\leq0.04$ for the constant-width injections and
$-1.03\leq\beta\leq-0.97$ for the injected $\nu^{-1}$ relation, each
mean having a standard error of approximately 0.015. The
lowest-frequency width is recovered to within 0.02~ms for 9 and 12
sub-bands and within 0.05~ms for 6 sub-bands. Within the adopted
exponential pulse-broadening model, neither the finite recording
window nor the width--tail covariance generates the observed
low-frequency excess.

\subsection{Dependence on the pulse-broadening function}
\label{app:pbf_tests}

To test the dependence of the recovered Gaussian-component width on
the assumed pulse-broadening function, we repeat the $\nu^{-1}$ width
injections using non-exponential responses
\citep{williamson1972,bhat2004}.
For each response, the recovered widths are averaged over 50
realisations, with the injected tail law normalised to
$\tau_{\rm ref}=7.98$~ms. The comparison uses the eight retained
sub-bands, weighted by their observed width uncertainties, and fits
one width normalisation, giving seven degrees of freedom.

The EMG-recovered widths from the thick-screen injections reproduce
the observed width evolution more closely than those from the
exponential-response injections, with $\chi^2=17.0$ and $46.7$,
respectively. The difference arises almost entirely from the
lowest-frequency sub-band. A uniform-medium response produces a
substantially steeper width increase, with $\chi^2=77$.

Because the observed width uncertainties do not represent the
full scatter about a single width--frequency relation, the absolute
$\chi^2$ values are inflated and should not be interpreted as formal
goodness-of-fit statistics. Their robust implication is the relative
ranking of the responses: thick screen, exponential response, and
uniform medium, in order of increasing $\chi^2$.

Fitting the simulated thick-screen profiles with an EMG also shifts an
injected $\alpha=-4$ to approximately $-3.5$, because the EMG absorbs
part of the rounded profile onset into the fitted Gaussian component.
These tests assess the dependence of the recovered parameters on the
assumed pulse-broadening function; they are not direct fits of the
alternative response functions to the observed burst profiles. We
therefore treat $\beta$ as a descriptive profile-shape index. The
measurements show an approximately $\nu^{-1}$ higher-frequency trend
and a low-frequency excess, but do not uniquely identify either with
the intrinsic emission duration.

\section{Scintillation search and injection tests}
\label{app:scint}

\subsection{Spectral autocorrelation search}
\label{app:acf_search}

The on-pulse spectrum in channel $i$ is
$S_i=\sum_{t\in W}[I_i(t)-B_i]$, where $W$ is the Catalog~2
emission window and $B_i$ is the off-pulse mean. We analyse four
contiguous runs of at least 300 unmasked channels:
600.0--610.7, 641.6--660.7, 664.3--693.2, and
700.7--714.4~MHz. These blocks contain 440, 784, 1186, and 561
channels, respectively.

For each block of $n$ channels, we initially form
$d_i^{(0)}=(S_i-\bar S)/\bar S$, where $\bar S$ is the mean of $S_i$ across the block, and fit the ACF with narrow and broad Lorentzian
components. The components are treated as inseparable if the fitted
broad width is less than four times the narrow width or if the narrow
width is uncertain by more than 50 per cent. This condition is met in
all four blocks. We therefore subtract a Gaussian-smoothed spectral
envelope $E_i$, using a kernel standard deviation of
$\max(5,0.1n)$ channels, and define
$d_i=(S_i-E_i)/\bar S$.

The one-sided ACF is
\begin{equation}
\mathrm{ACF}(k)
=
\frac{1}{n-k}
\sum_{i=1}^{n-k} d_i d_{i+k},
\label{eq:scint_acf}
\end{equation}
and is evaluated to a maximum lag of 2~MHz, following
\citet{masui2015}. Because each block is a contiguous run of unmasked
channels, every lag $k$ contains $n-k$ valid channel pairs. The ACF is
not divided by $\mathrm{ACF}(0)$, and zero lag is excluded because it
contains the radiometer-noise contribution.

We fit the non-zero-lag ACF with
$m^2/[1+(\delta\nu/\Delta\nu_{\rm d})^2]$, where $\delta\nu$ is the
frequency lag, $m$ is the rms fractional modulation index, and
$\Delta\nu_{\rm d}$ is the decorrelation bandwidth. With this
normalisation, $m^2$ is the extrapolated zero-lag amplitude of the
narrow ACF component.

The uncertainty assigned to each lag is the largest of the propagated
radiometer uncertainty \citep{jahnsschindler2026}, the scatter of an
equal-duration off-pulse control, and $N_{\rm s}^{-1/2}$, all expressed
in the same fractional-ACF normalisation. We estimate
\[
N_{\rm s}
=
\max\left[
\frac{n\Delta\nu_{\rm ch}}
     {\max(\Delta\nu_{\rm d},\Delta\nu_{\rm ch})},
\,2
\right],
\]
where $\Delta\nu_{\rm ch}=24.4$~kHz and $\Delta\nu_{\rm d}$ is the
fitted narrow-component bandwidth.

Parameter uncertainties are obtained from the
\texttt{curve\_fit} covariance matrix, treating the ACF lags as
independent. We classify a block as a detection only if
$m/\sigma_m>3$,
$\Delta\nu_{\rm d}/\sigma_{\Delta\nu_{\rm d}}>3$, and
$\Delta\nu_{\rm d}>1.5\Delta\nu_{\rm ch}$. Here,
$\sigma_m=\sigma_{m^2}/(2m)$, so the first requirement is equivalent
to $m^2/\sigma_{m^2}>1.5$.

No block meets all three criteria. The fitted modulation indices span
$m=0.21$--$0.42$, while the fitted bandwidths are
119--232~kHz with uncertainties of 107--225~kHz; no bandwidth differs
from zero by more than $1.2\sigma$. Because $m$ is constrained to be
non-negative, we do not interpret an average of the fitted modulation
indices as a physical measurement.

The on-pulse spectrum shows no consistent periodicity at the
16-channel CHIME upchannelisation period. At lags of 24--195~kHz, the
mean on-pulse ACF is 0.03--0.14, compared with 0.002--0.012 in an
equal-duration off-pulse control. The excess is clear in the 679- and
708-MHz blocks but not in the two lower-frequency blocks. These weak
correlations cannot be identified as scintillation because their
characteristic bandwidth is not significantly measured.

\subsection{Injection--recovery calibration}
\label{app:scint_injections}

To calibrate the search sensitivity, we generate a complex field on a
frequency grid 32 times finer than the native channel width using an
exponential delay profile with timescale
$(2\pi\Delta\nu_{\rm d})^{-1}$. This produces a Lorentzian intensity
ACF with half-width $\Delta\nu_{\rm d}$ and a unit-mean intensity $I$
with exponential statistics.

The scintillation gain is $G=1+m(I-1)$ and is averaged into the
native 24.4-kHz channels. The quoted injected $m$ is defined before
this averaging. We add $(G-1)$ times the Catalog~2 fitburst model of
the full burst dynamic spectrum to the observed data, thereby
retaining the measured noise, channel mask, and unmodelled spectral
structure.

We perform 100 trials for each
$m\in\{0.1,0.2,\ldots,0.8,1.0\}$ and for bandwidths spanning
4--150~kHz, including the NE2001 and YMW16 predictions at each block
centre. A trial is counted as a resolved recovery when at least one
block satisfies the detection criteria and its fitted
$\Delta\nu_{\rm d}$ lies within a factor of two of the injected value.
This bandwidth-matching condition prevents the burst's own spectral
structure, which produces formal detections at unrelated widths in
14--35 per cent of trials with $m\leq0.2$, from being counted as a
recovery.

NE2001-scale injections are recovered with the correct bandwidth in
11 per cent of trials at $m=0.3$, 66 per cent at $m=0.4$, and at
least 97 per cent for $m\geq0.5$. More generally, bandwidths of
50--150~kHz are recovered in at least 92 per cent of trials for
$m\geq0.6$.

Bandwidths of 18~kHz or less, including the YMW16 predictions, are
never recovered with a resolved bandwidth. At high modulation they
instead produce an unresolved, channel-limited signature: the formal
detection criteria are satisfied, but the fitted bandwidth does not
match the injected value. Such a signature appears in at least
96 per cent of trials for $m\geq0.7$ and in 47 per cent for
$m=0.5$. The present search therefore cannot measure a YMW16-scale
decorrelation bandwidth, although strong modulation on this scale is
disfavoured.

For resolved bandwidths, the recovered modulation index exceeds the
injected value by approximately 0.1. We therefore use the injections
to calibrate detection sensitivity rather than to infer an unbiased
physical modulation index. These tests assume box-car channel
responses and a constant $\Delta\nu_{\rm d}$ within each frequency
block.

\FloatBarrier

\section*{Data availability}

The Stokes-$I$ dynamic spectrum of FRB~20210415C is publicly available
in the Second CHIME/FRB Catalog data release \citep{chime2026}
(\texttt{https://doi.org/10.11570/25.0066}). The analysis code and the
scripts that reproduce all results and figures in this paper are available at \texttt{https://github.com/popv7/frb20210415c-scattering}.

\section*{Acknowledgements}

We thank the CHIME/FRB Collaboration for the public release of the
Second CHIME/FRB Catalog. F.R.'s research is supported by the NSERC
Discovery Grant RGPIN-2024-06346.

\section*{Declaration of generative AI and AI-assisted technologies in the manuscript preparation process}

During the preparation of this work, the authors used OpenAI's ChatGPT
to assist with manuscript organisation and language refinement, and
Anthropic's Claude to assist with code refinement, execution of
analysis scripts, figure generation, and drafting text revisions. The
authors reviewed and edited all AI-assisted outputs and take full
responsibility for the content of the manuscript.

\bibliographystyle{elsarticle-harv}
\bibliography{refs}

\end{document}